\RequirePackage{lineno}
\documentclass[aps,floats]{revtex4}  
\usepackage{graphicx}  
\usepackage{dcolumn}   
\usepackage{bm}        
\usepackage{amssymb}   
\usepackage{rotating}  
\usepackage{afterpage}
\def\MET{{\mbox{$E\kern-0.57em\raise0.19ex\hbox{/}_{T}$}}}
\def\met{{\mbox{$E\kern-0.57em\raise0.19ex\hbox{/}_{T}$}}}
\def\DZ{D0}

\def\pp{$p\bar{p}$}

\def\ttbar{$t\bar{t}$}

\def\lmet{$WH\rightarrow \ell\kern-0.45em\raise0.19ex\hbox{/} \nu b\bar{b}$}

\def\zee{\ensuremath{Z\rightarrow e^+e^-}}
\def\zmm{\ensuremath{Z\rightarrow\mu^+\mu^-}}
\def\wlnu{\ensuremath{W\rightarrow \ell\nu}}

\def\tanb{\ensuremath{\mathrm{tan}\beta}}
\def\etau{\ensuremath{\tau_e\tau_{\mathrm{had}}}}
\def\mutau{\ensuremath{\tau_\mu\tau_{\mathrm{had}}}}
\def\emu{\ensuremath{\tau_e\tau_\mu}}
\def\ztt{\ensuremath{Z\rightarrow\tau\tau}}

\def\Att{\ensuremath{A\rightarrow\tau\tau}}
\def\zll{\ensuremath{\mathrm{Z}/\gamma^*\rightarrow \ell^+\ell^-}}
\RequirePackage{xspace}
\newcommand{\jprlBase}       {Phys.\ Rev.\ Lett.\xspace}
\newcommand{\jprl}      [1]  {\jprlBase\ {\bf #1}}

\begin{document}
\rightline{FERMILAB-FN-0851-E}
\rightline{CDF Note 10099}
\rightline{\DZ\ Note 6036-CONF}
\vskip0.5in

\title{Combined CDF and \DZ\ upper limits on MSSM Higgs boson production in tau-tau final states with up to 2.2 fb$^{-1}$ of data\\[2.5cm]}
\author{
The TEVNPH Working Group\footnote{The Tevatron
New Phenomena and Higgs working group can be contacted at
TEVNPHWG@fnal.gov. More information can be found at http://tevnphwg.fnal.gov/.}
 } 
\affiliation{\vskip0.3cm for the CDF and \DZ\ Collaborations\\ \vskip0.2cm
\today} 
\begin{abstract}
Combined results are presented on the search for a neutral Higgs boson in the di-tau final state using 1.8~fb$^{-1}$ and 2.2~fb$^{-1}$ of integrated luminosity collected at the CDF and \DZ\ experiments respectively. Data were collected in $p\bar{p}$ collisions at a centre of mass energy of 1.96~TeV during RunII of the Tevatron. Limits are set on the cross section $\times$ branching ratio ranging from 13.6~pb to 0.653~pb for Higgs masses from 90~GeV to 200~GeV respectively. The results are then interpreted as limits in four different benchmark scenarios within the framework of the MSSM.
\vskip0.3in 
{\hspace*{5.5cm}\em Preliminary Results}
\end{abstract}

\maketitle

\newpage
\section{Introduction} 

Spontaneous symmetry breaking in the electroweak sector is an attractive solution to the problem of the origin of particle masses within the Standard Model (SM). 
However, extreme fine tuning is required to avoid divergencies in radiative corrections to the Higgs mass. 
Supersymmetry (SUSY) as an extension to the SM, provides a natural means to avoid this as well as potentially providing a candidate for dark matter and GUT-scale unification. The Minimal Supersymmetric Standard Model (MSSM) \cite{mssm} requires the introduction of two Higgs doublets and predicts the existence of five physical Higgs bosons after symmetry breaking: three neutral ($h$, $H$, and $A$) and two charged ($H^{\pm}$). The ratio of the vacuum expectation values of the two doublets is denoted by \tanb. For high values of \tanb\ two of the three neutral Higgs bosons have approximately the same mass and couplings. These couplings are enhanced with respect to the charged leptons and down-type quarks by a factor \tanb\ relative to the SM, and suppressed for the neutrinos and up-type quarks. The near degeneracy contributes an additional factor two enhancement in the cross section. Thus for low $M_{A}$ and high \tanb\ the Tevatron can probe a number of benchmark scenarios in the MSSM complementing the regions of the SUSY parameter space probed by the LEP experiments\cite{LEP_exclu}.

The results presented here represent an update to the previous combination \cite{combo}. The same inputs, with an additional mass hypothesis at 90 GeV,  are used but the treatment of correlations between the systematic uncertainties on those backgrounds estimated from Monte-Carlo simulations has been improved and the intepretation of the results in the MSSM uses calculations from the latest version of {\sc feynhiggs}\cite{feynhiggs}. 

\section{Analysis Summary}

The CDF and \DZ\ detectors are described in detail elsewhere \cite{cdf,dzero}. The searches combined here are described in detail in \cite{cdfhtt,p17htt,p20htt} and earlier published results from CDF and \DZ\ can be found in \cite{p14htt,cdfhttpub}.  Searches are performed at CDF and \DZ\ for MSSM Higgs boson production with subsequent decays to taus in a number of channels characterised by the decay products of the $\tau$ leptons. Included in this combination are 1.8~fb$^{-1}$ of data collected at CDF in three final states: \etau, \mutau\ and \emu, (where $\tau_e, \tau_\mu,$ and $\tau_{\mathrm{had}}$ denote $\tau$ decays to electron, muon and hadrons respectively) and 1.0~fb$^{-1}$ in the same three channels and an additional 1.2~fb$^{-1}$ in the \mutau\ final state collected at \DZ. Additionally, the searches from \DZ\ are split further depending on the hadronic decay multiplicity.

\subsection{Lepton Identification}
Electrons are identified through their characteristic energy deposits in the calorimeters. Reconstructed clusters of energy in the calorimeters are required to be isolated and match a reconstructed track, suppressing photon backgrounds. Muons are identified by matching charged tracks in the central tracking detectors with hits in the muon detectors. Muon candidates are also required to be isolated in both the central tracking detectors and in the calorimetry. 

Hadronic decays of $\tau$ leptons are identified at CDF by selecting isolated narrow clusters in the calorimeter with 1 or 3 spatially matched charged tracks. These are reconstructed using a variable sized cone algorithm whose angle, $\alpha$, is set to be the minimum of 10$^\circ$ and $\left( 5 ~\mathrm{GeV}\right)/E_{cl}$ radians, where $E_{cl}$ is the calorimeter cluster energy. 
Strict isolation limits on the number of tracks and the calorimeter energy within an annulus around the candidate from $\alpha$ out to an angle of 30$^\circ$ are used to suppress quark and gluon jets. In the case of three-prong candidates the sum of the charges of the tracks is required to be $\pm$1. One-prong candidates are rejected if found to be consistent with an electron having undergone significant bremsstrahlung. 

In the \DZ\ analyses, the hadronic decays of the $\tau$ are divided into three categories: $\tau$ types 1 and 2 are one-prong candidates with energy either in only the hadron calorimeter ($\pi^\pm$ like) or in both the electromagnetic and hadron calorimeters ($\rho^\pm$  like) respectively; $\tau$ type 3 is a three-prong candidate with an invariant mass (constructed from the three tracks) below $1.7$~GeV and matching energy deposits in the calorimeters. A neural network (NN) is trained for each type to separate hadronic tau decays from jets using MC \ztt\ as the signal and multi-jet events taken from data as the background. An additional NN is trained on electron Monte-Carlo events and is employed to reduce backgrounds from electrons faking type 2 taus.

\subsection{Signal, Backgrounds and Event Selection}
The acceptance for signal is determined from Monte-Carlo simulations, using the {\sc pythia}\cite{pythia} event generator with {\sc cteq5l} (CDF) and {\sc cteq6L} \cite{cteq6l} (D0) parton sets and {\sc tauola}\cite{tauola} to simulate the decays of the final state $\tau$-leptons. The response of the detectors is modeled using {\sc geant}\cite{geant} based simulations. Two production modes, $gg\rightarrow A$ and $b\bar{b}\rightarrow A$ are considered by CDF, whereas at \DZ\ only $gg\rightarrow A$ is simulated - the acceptances are seen to be very similar for both production modes. In the interpretation of the results in the framework of the MSSM as limits in the \tanb-$M_A$ plane both production modes are taken into account as well as an additional factor of approximately two on the cross section due to the near degeneracy of two of the three neutral Higgs bosons.
  Most Standard Model backgrounds have been generated with {\sc pythia}: \zll, \wlnu, di-boson production, \ttbar\ ({\sc comphep + pythia})\cite{comphep}. $W$ and $Z$ boson samples where there is one or more additional jets in the final state have been simulated with {\sc alpgen}\cite{alpgen} with matching to {\sc pythia} for hadronization. Di-boson and \ttbar\ samples are normalised using calculations to next-to-LO (NLO)\cite{ttbar,diboson} while $Z/\gamma^*$ samples are generally normalised to next-to-NLO (NNLO) \cite{znnlo}.

Events are selected by the trigger using inclusive electron and muon (D0) and lepton plus track (CDF) triggers and after offline reconstruction candidate events must contain two isolated opposite charged final state leptons ($e,\mu, \tau$).
Leading sources of background are: $Z/\gamma^*\rightarrow\tau\tau$, multi-jet, $W\rightarrow e\nu, \mu\nu, \tau\nu$, $Z\rightarrow\mu\mu, Z\rightarrow$ee, di-boson ($WW,WZ,ZZ,W\gamma,Z\gamma$), and $t\bar{t}$-pair production. 
In the \emu\ channel at CDF, events from the sidebands of the lepton isolation are used to determine the jet backgrounds. For the \etau\ and \mutau\ channels the jet backgrounds where a jet fakes a $\tau_{\mathrm{had}}$ are estimated by weighting data events passing very loose cuts with the jet-$\tau$ fake probability measured in an independent jet sample.
The multi-jet contribution from data collected at \DZ\ is estimated using either \etau\ candidate events where the electron and $\tau$ have the same charge or using inverted lepton selection criteria (\mutau\ and \emu\ channels). The normalisation of the $W$ production backgrounds is estimated from a data sample dominated by $W+$jet events. 

In the \etau\ and \mutau\ channels the electron or muon are required to be isolated and have a transverse momentum, $p_T > 10$ (CDF) or 15 (D0) GeV. One-prong hadronic tau candidates are accepted with $p_T > 15$ GeV (CDF), $16.5$~GeV (D0) and three-prong are required to have $p_T > 20$~GeV (CDF) $22$~GeV (D0). Additional cuts are placed on the scalar sum of transverse momenta in the event at CDF, $H_T = |p_T^{e/\mu}| + |p_T{\tau_{\mathrm{had}}}| + |\met| > 55$~GeV, where \met is the momentum imbalance in the transverse plane. In one-prong events where the rate at which jets fake taus is lower a slightly looser cut is used, $H_T > 50 $ or $45$~GeV for \etau\ and \mutau\ respectively. Further cuts on the relative directions of the taus and the \met (CDF and D0) and the transverse mass (D0) $M_{T} = \sqrt{2p^{e/\mu}_T\met(1-\cos\Delta\phi)}$, where $\Delta\phi$ is the azimuthal angle between the electron or muon and the hadronic tau, serve to suppress background contributions from $W$+jets production. 

In the CDF \emu\ channel events are selected requiring one central electron and one central muon with: $\mathrm{min}(E_T^e,p^\mu_T) > 6~\mathrm{GeV}, \mathrm{max}(E_T^{e},p_T^{\mu}) > 10 ~\mathrm{GeV},$ and $|E_T^e| + |p^\mu_T| > 30$~GeV. \DZ\ make a similar selection, where: $p_T^{\mu} > 10$~GeV and $p_T^e > 12$~GeV and the invariant mass of the electron-muon pair exceeds $20$~GeV and  $|E_T^e| + |p^\mu_T| + |\met| > 65$~GeV. Table \ref{tab:events} shows the expected number of backgrounds, observed events in data and the signal efficiency for $M_A = 130$~GeV.

\begin{figure}[!htbp]
\begin{center}
\begin{tabular}{cc}
 \includegraphics[width=0.45\textwidth]{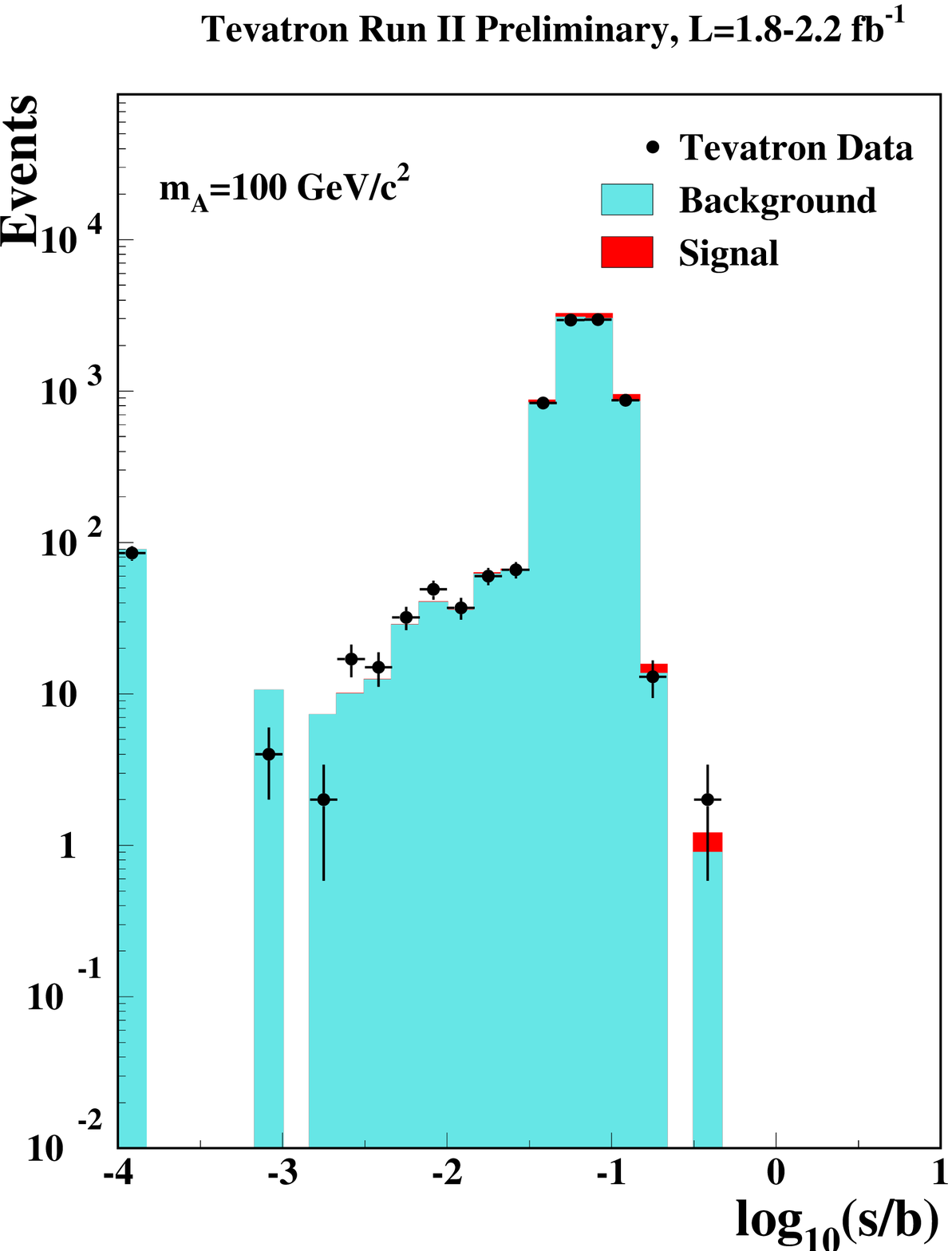} &
 \includegraphics[width=0.45\textwidth]{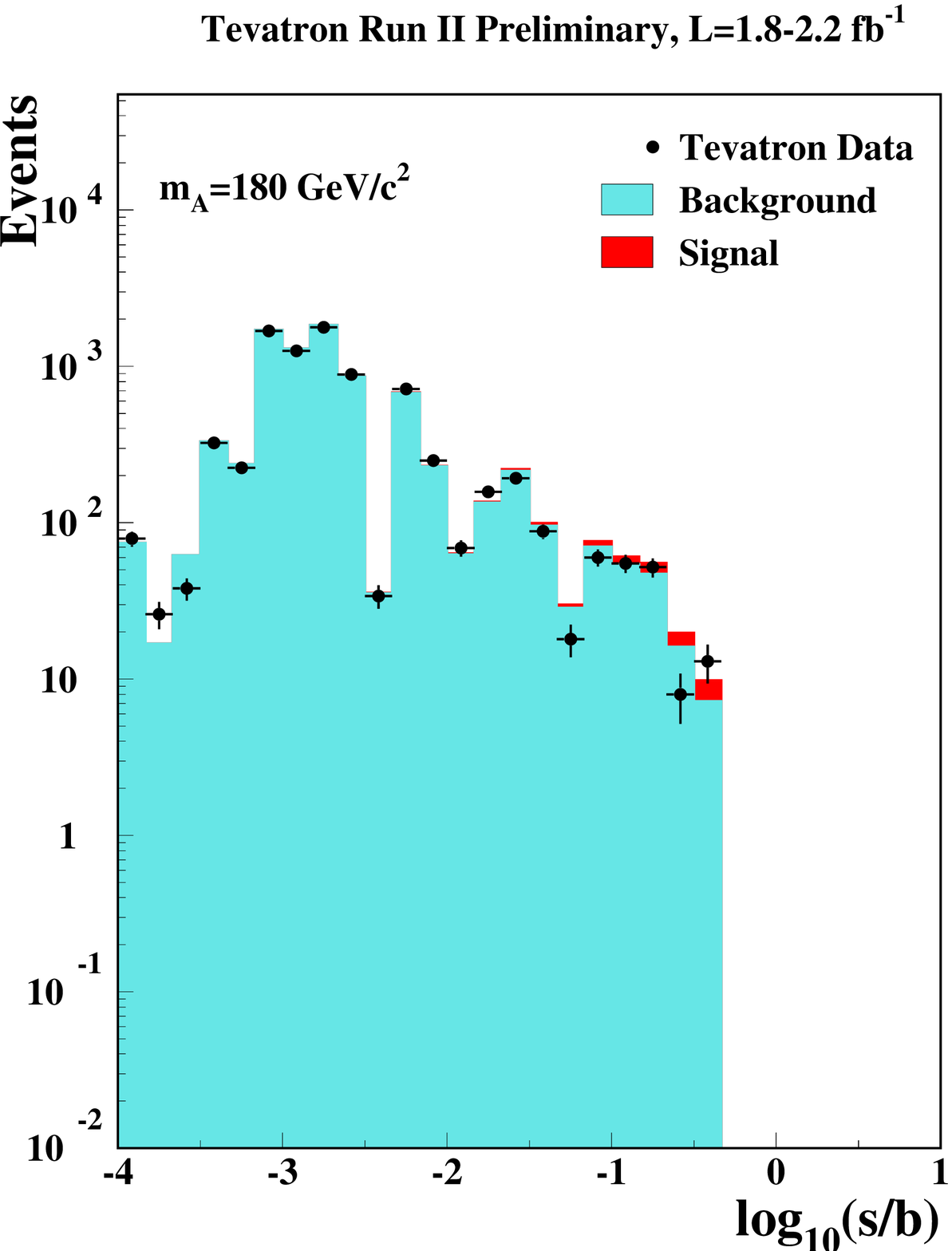}\\
\end{tabular}
\end{center}
\caption{Events binned by the ratio of expected signal to expected background for a signal of $M_A=100$~GeV, and $\sigma\times\mathrm{Br}=2.0$pb (left) and $M_A=180$~GeV, and $\sigma\times\mathrm{Br}=0.66$pb (right) \label{fig:logsoverb}.}
\end{figure}

\begin{figure}[!htbp]
\begin{center}
\begin{tabular}{cc}
 \includegraphics[width=0.45\textwidth]{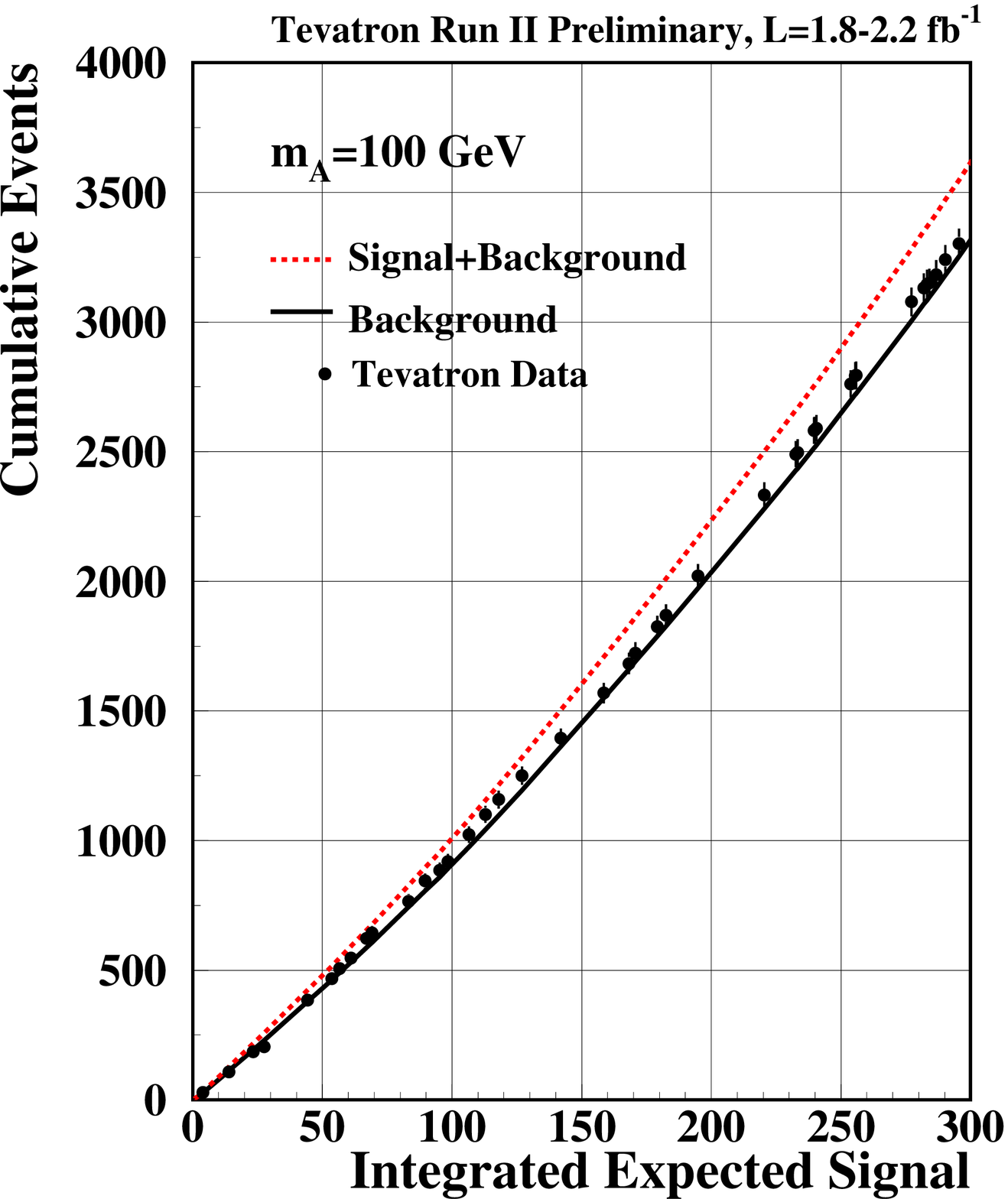} &
 \includegraphics[width=0.45\textwidth]{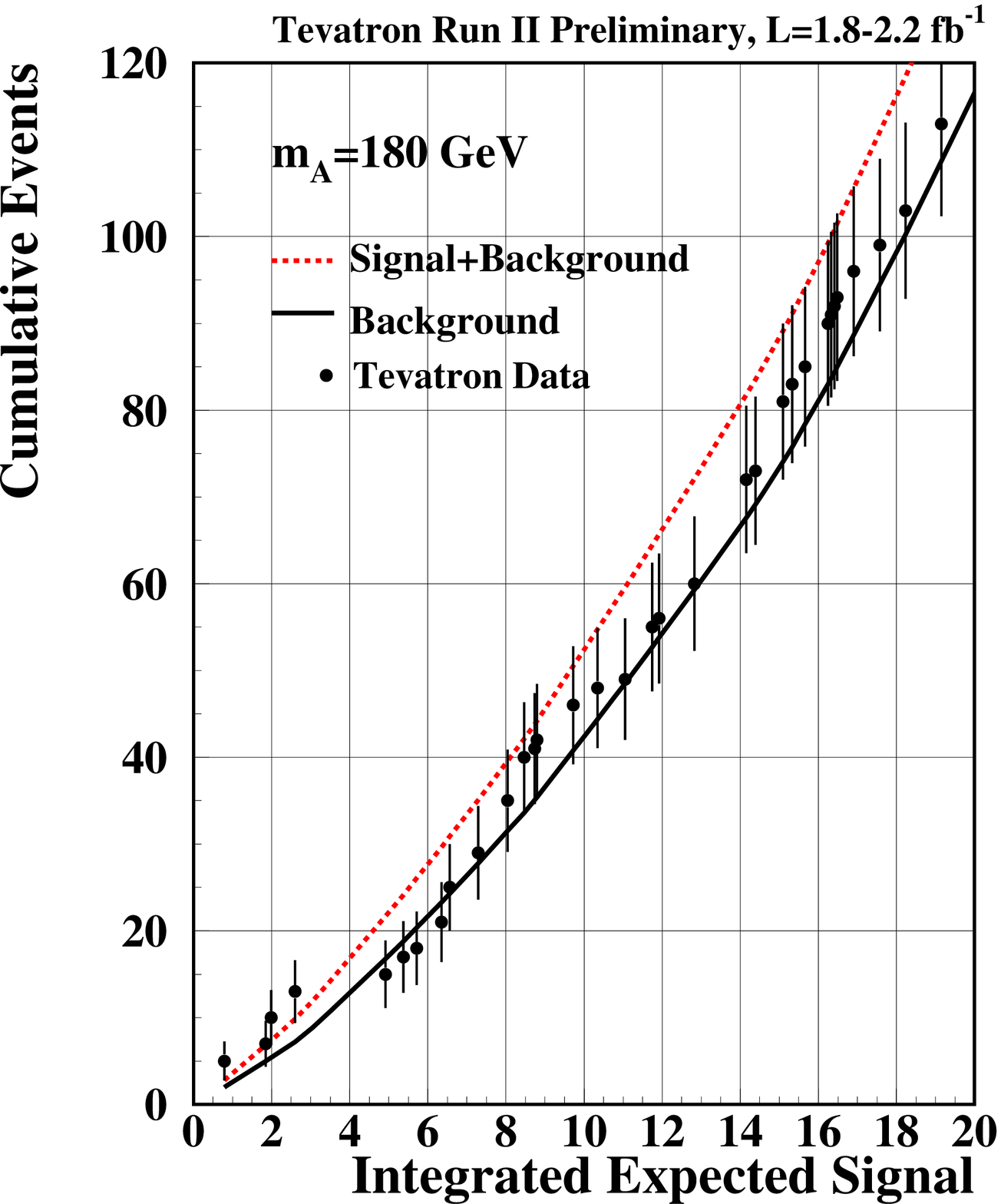}\\
\end{tabular}
\end{center}
\caption{Integrated distributions of S/B, starting at the high S/B side for $M_A = 100$~GeV,  $\sigma\times $Br$ = 2.0$pb (left) and $M_A=180$~GeV, $\sigma\times\mathrm{Br}$=0.66pb (right). The total signal+background and background-only integrals are shown separately with data superimposed. Data points are only plotted for those bins with data events.\label{fig:intsoverb}}
\end{figure}

In setting the limits, events from regions of phase space with a similar ratio of expected signal (S) to background (B) can be combined without loss of sensitivity. Thus a useful way to visualize the comparison of expected backgrounds and the observed data is to show the event distributions binned in this ratio S/B. For the channels combined in the results presented in this note these distributions are shown in Figure \ref{fig:logsoverb}. The left hand plot is for a signal, $M_A = 100$~GeV and $\sigma\times $Br$ = 2.0$~pb and the right hand plot for a signal of $M_A=180$~GeV and $\sigma\times\mathrm{Br}$=0.66~pb. Good agreement is observed between the data and expected backgrounds. The integrals of these distributions starting from the high S/B side and working downwards are shown in Figure \ref{fig:intsoverb}, displaying the signal+background, background-only and data sums. 

\begin{table}
\begin{center}
\begin{tabular}{l|ccc|ccc}
\hline
                     & \multicolumn{3}{c}{CDF} & \multicolumn{3}{c}{D\O\ }\\
Source               &   \emu\             & \etau\              & \mutau\             & \emu\                & \etau\               & \mutau\         \\ \hline
\ztt                 & 605 $\pm$ 51        & 1378 $\pm$ 117      &  1353 $\pm$ 116     & 212 $\pm$ 20         &   581 $\pm$ 5        &  2153 $\pm$ 156  \\
\zee /\zmm           & 19.4 $\pm$ 5.7      &  70 $\pm$ 10        &  107 $\pm$ 13       & 10.4 $\pm$ 1.3       &   31 $\pm$ 2         &   66 $\pm$ 8    \\
diboson + $t\bar{t}$ & 20.5 $\pm$ 7.0      &  8.2 $\pm$ 4.2      &  6.6 $\pm$ 3.7      & 6.1 $\pm$ 0.6        &   3.1 $\pm$ 0.3      &  16 $\pm$ 3 \\
multi-jet + \wlnu    &  57.1 $\pm$ 13.5    &  467 $\pm$ 73       &  285 $\pm$ 46       & 37.9 $\pm$ 7.7       &   374 $\pm$ 48       & 216 $\pm$ 41    \\ \hline
Total Background     & 702 $\pm$ 55        &  1922 $\pm$ 141     &  1752 $\pm$ 129     &  266 $\pm$ 22        &  989 $\pm$ 82        & 2451 $\pm$ 162      \\\hline
Data                 & 726                 &  1979               &  1666               &  274                 &  1034                & 2340            \\\hline
Signal Efficiency /\%& 0.32 $\pm$ 0.01     &  0.77 $\pm$ 0.01    &  0.67 $\pm$ 0.01    &  0.41 $\pm$ 0.03     & 0.73 $\pm$ 0.03      & 0.99 $\pm$ 0.05 \\\hline
\end{tabular}
\end{center}
\caption{Expected numbers of background and observed data events and signal efficiency for $M_A=130$ GeV. Errors include full systematic uncertainties, that are in some cases correlated.\label{tab:events}}
\end{table}

\section{Combination} 

To gain confidence that the final result does not depend on the
details of the statistical formulation, two types of combinations are performed, using the
Bayesian and  Modified Frequentist approaches, which give similar results
(within 10\%).
Both methods rely on distributions in the final discriminants, and not just on
their single integrated values.  Systematic uncertainties enter as uncertainties on the
expected number of signal and background events, as well
as on the distribution of the discriminants in 
each analysis (``shape uncertainties'').
Both methods use likelihood calculations based on Poisson
probabilities. In all channels the visible mass distribution is used to set limits. 
\subsection{Bayesian Method}

Because there is no experimental information on the production cross section for
the Higgs boson, in the Bayesian technique~\cite{CDFHiggs} a flat prior is assigned
for the total number of selected Higgs events.  For a given Higgs boson mass, the
combined likelihood is a product of likelihoods for the individual
channels, each of which is a product over histogram bins:

\begin{equation}
{\cal{L}}(R,{\vec{s}},{\vec{b}}|{\vec{n}},{\vec{\theta}})\times\pi({\vec{\theta}})
= \prod_{i=1}^{N_C}\prod_{j=1}^{Nbins} \mu_{ij}^{n_{ij}} e^{-\mu_{ij}}/n_{ij}!
\times\prod_{k=1}^{n_{np}}e^{-\theta_k^2/2}
\end{equation}

\noindent where the first product is over the number of channels
($N_C$), and the second product is over histogram bins containing
$n_{ij}$ events, binned in  ranges of the final discriminants used for
individual analyses, such as the di-jet mass, neural-network outputs, 
or matrix-element likelihoods.
 The parameters that contribute to the
expected bin contents are $\mu_{ij} =R \times s_{ij}({\vec{\theta}}) + b_{ij}({\vec{\theta}})$ 
for the
channel $i$ and the histogram bin $j$, where $s_{ij}$ and $b_{ij}$ 
represent the expected background and signal in the bin, and $R$ is a scaling factor
applied to the signal to test the sensitivity level of the experiment.  
Truncated Gaussian priors are used for each of the nuisance parameters                                               
$\theta_k$, which define
the
sensitivity of the predicted signal and background estimates to systematic uncertainties.
These
can take the form of uncertainties on overall rates, as well as the shapes of the distributions
used for combination.   These systematic uncertainties can be far larger
than the expected Higgs signal, and are therefore important in the calculation of limits. 
The truncation
is applied so that no prediction of any signal or background in any bin is negative.
The posterior density function is
then integrated over all parameters (including correlations) except for $R$,
and a 95\% credibility level upper limit on $R$ is estimated
by calculating the value of $R$ that corresponds to 95\% of the area
of the resulting distribution.

\subsection{Modified Frequentist Method}

The Modified Frequentist technique relies on the $CL_s$ method, using
a log-likelihood ratio (LLR) as test statistic~\cite{DZHiggs}:
\begin{equation}
LLR = -2\ln\frac{p({\mathrm{data}}|H_1)}{p({\mathrm{data}}|H_0)},
\end{equation}
where $H_1$ denotes the test hypothesis, which admits the presence of
SM backgrounds and a Higgs boson signal, while $H_0$ is the null
hypothesis, for only SM backgrounds.  The probabilities $p$ are
computed using the best-fit values of the nuisance parameters for each
event, separately for each of the two hypotheses, and include the
Poisson probabilities of observing the data multiplied by Gaussian
constraints for the values of the nuisance parameters.  This technique
extends the LEP procedure which does not involve a
fit, in order to yield better sensitivity when expected signals are
small and systematic uncertainties on backgrounds are
large~\cite{collie}.

The $CL_s$ technique involves computing two $p$-values, $CL_{s+b}$ and $CL_b$.
The latter is defined by
\begin{equation}
1-CL_b = p(LLR\le LLR_{\mathrm{obs}} | H_0),
\end{equation}
where $LLR_{\mathrm{obs}}$ is the value of the test statistic computed for the
data. $1-CL_b$ is the probability of observing a signal-plus-background-like outcome 
without the presence of signal, i.e. the probability
that an upward fluctuation of the background provides  a signal-plus-background-like
response as observed in data.
The other $p$-value is defined by
\begin{equation}
CL_{s+b} = p(LLR\ge LLR_{\mathrm{obs}} | H_1),
\end{equation}
and this corresponds to the probability of a downward fluctuation of the sum
of signal and background in 
the data.  A small value of $CL_{s+b}$ reflects inconsistency with  $H_1$.
It is also possible to have a downward fluctuation in data even in the absence of
any signal, and a small value of $CL_{s+b}$ is possible even if the expected signal is
so small that it cannot be tested with the experiment.  To eliminate the possibility
of  excluding  a signal to which there is insufficient sensitivity 
(an outcome  expected 5\% of the time at the 95\% C.L., for full coverage),
we use the quantity $CL_s=CL_{s+b}/CL_b$.  If $CL_s<0.05$ for a particular choice
of $H_1$, that hypothesis is deemed excluded at the 95\% C.L.

Systematic uncertainties are included  by fluctuating the predictions for
signal and background rates in each bin of each histogram in a correlated way when computing $CL_{s+b}$ and $CL_b$.

\subsection{Systematic Uncertainties} 
The uncertainty on the measurement of the integrated luminosity is 5.8\%
(CDF) and 6.1\% (D0).  
Of this value, 4\% arises from the uncertainty
on the inelastic \pp~scattering cross section, which is correlated
between CDF and D0. 
The uncertainty on the rates for \ttbar\ production and for single and di-electroweak boson production are taken as correlated between the two
experiments. As the methods of measuring the multi-jet (``QCD'')
backgrounds differ between CDF and D0, there is no
correlation assumed between these rates.  
The calibrations of fake leptons, unvetoed $\gamma\rightarrow e^+e^-$ conversions,
$b$-tag efficiencies and mistag rates are performed by each collaboration
using independent data samples and methods, hence are considered uncorrelated.

Tables \ref{tab:systcdfetau} to \ref{tab:systdzemu} summarize the various contributions to the systematics uncertainties to the input distributions used in the limit setting, broken down by experiment and channel. Entries in the tables labeled as ``shape'' systematics do not have the same value across all bins of the relevant distribution and model the systematic variation of the shape for that source of uncertainty. In these cases the number given is the event weighted mean fluctuation away from the nominal distribution - i.e. related to the flat component of the uncertainty.

\begin{table}[!htbp]
\begin{center}
\begin{tabular}{r|cccccc}
\hline
       Contribution & Signal                & \zee                   & \ztt               & $t\bar{t}$                 & diboson               & QCD                   \\ \hline
          Jet energy scale (shape) &                   0.12& $^{+  0.30}_{+  0.22}$& $^{+  0.05}_{   0.00}$&                   0.56&                   0.73&                    0.0\\
          Electron identification&                    2.4&                    2.4&                    2.4&                    2.4&                    2.4&                    0.0\\
          Electron energy scale (shape) & $^{+  0.32}_{  -0.23}$& $^{   0.00}_{+  0.30}$& $^{+  0.77}_{  -0.50}$& $^{  -0.19}_{+  0.28}$& $^{  -0.09}_{+  0.16}$&                    0.0\\
          Tau identification&                    4.2&                    4.2&                    4.2&                    4.2&                    4.2&                    0.0\\                        
          Tau energy scale (shape) & $^{+  0.29}_{  -0.22}$& $^{+  0.22}_{+  0.23}$& $^{+  0.82}_{  -0.63}$& $^{+  0.42}_{  -0.75}$& $^{+  0.63}_{  -0.31}$&                    0.0\\
          $gg\rightarrow A$ acceptance&                    2.1&                    0.0&                    0.0&                    0.0&                    0.0&                    0.0\\
          $bb\rightarrow A$ acceptance&                    3.6&                    0.0&                    0.0&                    0.0&                    0.0&                    0.0\\
         MC Cross sections&                    0.0&                    2.2&                    2.2&                    10.0&                    6.0&                    0.0\\
        QCD&                    0.0&                    0.0&                    0.0&                    0.0&                    0.0&                   15.0\\
         Luminosity&                    5.8&                    0.0&                    5.8&                    5.8&                    5.8&                    0.0\\\hline
\end{tabular}
\end{center}
\caption{Percentage systematic uncertainties for each distribution in the CDF \etau\ analysis. Signal uncertainties are for $M_A = 130$ GeV.\label{tab:systcdfetau}}
\end{table}

\begin{table}[!htbp]
\begin{center}
\begin{tabular}{r|cccccc}
\hline
       Contribution & Signal                & \zmm                 & \ztt               & $t\bar{t}$                 & diboson               & QCD                   \\ \hline
              Jet energy scale (shape) &                   0.07& $^{+  0.24}_{  -0.38}$&                    0.0& $^{+  0.54}_{  -0.48}$& $^{+  0.46}_{  -0.58}$&                    0.0\\
          Muon identification&                    2.7&                    2.7&                    2.7&                    2.7&                    2.7&                    0.0\\
          Tau identification&                    4.2&                    4.2&                    4.2&                    4.2&                    4.2&                    0.0\\
          Tau energy scale (shape) & $^{+  0.23}_{  -0.08}$&                    0.0& $^{+  0.54}_{  -0.77}$& $^{+  0.97}_{  -0.75}$& $^{+  0.40}_{  -0.70}$&                    0.0\\
          $gg\rightarrow A$ acceptance&                    2.1&                    0.0&                    0.0&                    0.0&                    0.0&                    0.0\\
          $bb\rightarrow A$ acceptance&                    3.6&                    0.0&                    0.0&                    0.0&                    0.0&                    0.0\\
         MC cross sections&                    0.0&                    2.2&                    2.2&                    10.0&                    6.0&                    0.0\\
        QCD&                    0.0&                    0.0&                    0.0&                    0.0&                    0.0&                   20.0\\
         Luminosity&                    5.8&                    0.0&                    5.8&                    5.8&                    5.8&                    0.0\\ \hline
\end{tabular}
\end{center}
\caption{Percentage systematic uncertainties for each distribution in the CDF \mutau\ analysis. Signal uncertainties are for $M_A = 130$ GeV.\label{tab:systcdfmutau}}

\end{table}

\begin{table}[!htbp]
\begin{center}
\begin{tabular}{r|ccccccc}
\hline
       Contribution           & Signal             & \zee   & \zmm                 & \ztt               & $t\bar{t}$                 & diboson               & QCD                   \\ \hline
Electron energy scale (shape) & $^{+  0.23}_{  -0.46}$&    0.0& $^{  -0.36}_{  -0.60}$  & $^{+  0.72}_{  -0.62}$  & $^{   0.00}_{+  0.26}$   &     0.0&            0.0\\
Jet energy scale (shape)      & $^{  -0.08}_{   0.00}$&    0.0& $^{  -0.34}_{  -0.30}$  & $^{  -0.05}_{   0.00}$  &                    0.57&    0.29&                    0.0\\
Electron identification       &                 2.4&    2.4&                    2.4&                    2.4&                    2.4&    2.4&                    0.0\\
Muon identification           &                 2.7&    2.7&                    2.7&                    2.7&                    2.7&    2.7&                    0.0\\
$gg\rightarrow A$ acceptance  &                 2.1&    0.0&                    0.0&                    0.0&                    0.0&    0.0&                    0.0\\
$bb\rightarrow A$ acceptance  &                 3.6&    0.0&                    0.0&                    0.0&                    0.0&    0.0&                    0.0\\
MC Cross sections             &                 0.0&    2.2&                    2.2&                    2.2&                    10.0&   6.0&                    0.0\\
QCD                           &                 0.0&    0.0&                    0.0&                    0.0&                    0.0&    0.0&                   20.0\\
Luminosity                    &                 5.8&    5.8&                    5.8&                    5.8&                    5.8&    5.8&                    0.0\\ \hline
\end{tabular}
\end{center}
\caption{Percentage systematic uncertainties for each distribution in the CDF \emu\ analysis. Signal uncertainties are for $M_A = 130 $ GeV.\label{tab:systcdfemu}}
\end{table}

\begin{table}[!htbp]
\begin{center}
\begin{tabular}{r|ccccccc}
\hline
       Contribution & Signal           & diboson               & QCD                   & \ensuremath{t\bar{t}} & \wlnu                   & \zee              & \ztt \\ \hline
Electron Identification&          3.3&                    3.3&                    0.0&                    3.3&                    3.3&                  3.3&               3.3\\
Electron-tau fake rate&          0.0&                    0.0&                    0.0&                    0.0&                    0.0&                   13 &              0.0\\
Tau identification&               6.0&                    5.3&                    0.0&                    7.1&                    5.6&                  3.9&               4.1\\
Tau track reconstruction    &              1.0&                    1.0&                    0.0&                    1.0&                    1.0&                  1.0&               1.0\\
Tau energy scale (shape)  &       0.4&                    0.0&                    0.0&                    0.0&                    0.0&                  0.0&               1.3\\
Trigger (shape)   &               3.8&                    4.1&                    0.0&                    3.0&                    4.4&                  4.2&               5.9\\
Signal acceptance               &               4.0&                    0.0&                    0.0&                    0.0&                    0.0&                  0.0&               0.0\\
MC cross sections &               0.0&                    5.0&                    0.0&                    5.0&                    0.0&                  5.0 &               5.0\\
W+jets            &               0.0&                    0.0&                    0.0&                    0.0&                    6.8&                  0.0&               0.0\\
QCD               &               0.0&                    0.0&                   13.0&                    0.0&                    0.0&                  0.0&               0.0\\
Luminosity        &               6.1&                    6.1&                    0.0&                    6.1&                    6.1&                  6.1&               6.1\\\hline
\end{tabular}
\end{center}
\caption{Percentage systematic uncertainties for each distribution in the D\O\ \etau\ analysis - combined across all three tau categories. Signal uncertainties are for $M_A = 130$ GeV. \label{tab:systdzetau}}
\end{table}

\begin{table}[!htbp]
\begin{center}
\begin{tabular}{r|ccccccc}
\hline
       Contribution & Signal                & diboson               & QCD                   & \ensuremath{t\bar{t}}                    & \wlnu                   & \zmm                 & \ztt               \\ \hline
Muon identification &                 1.1&                    1.1&                    0.0&                    1.1&                    1.1&                    1.1&                    1.1\\
Tau identification&                 4.2&                    3.9&                    0.0&                    4.2&                    5.6&                    3.9&                    3.9\\
Tau track reconstruction   &         1.0&                    1.0&                    0.0&                    1.0&                    1.0&                    1.0&                    1.0\\
Tau energy scale (shape) &                 0.79&                   0.0&                    0.0&                    0.0&                    0.0&                    0.0&                    1.3\\
Trigger          &                    3.0&                    3.0&                    0.0&                    3.0&                    3.0&                    3.0&                    3.0\\
Signal acceptance              &                    4.0&                    0.0&                    0.0&                    0.0&                    0.0&                    0.0&                    0.0\\
MC cross sections&                    0.0&                    5.0&                    0.0&                    5.0&                    0.0&                    5.0&                    5.0\\
W+jets           &                    0.0&                    0.0&                    0.0&                    0.0&                   13.0&                    0.0&                    0.0\\
QCD              &                    0.0&                    0.0&                     32&                    0.0&                    0.0&                    0.0&                    0.0\\
Luminosity       &                    6.1&                    6.1&                    0.0&                    6.1&                    6.1&                    6.1&                    6.1\\\hline
\end{tabular}
\end{center}
\caption{Percentage systematic uncertainties for each distribution in the  D\O\ \mutau\ - (RunIIa) - combined across all three tau categories. Signal uncertainties are for $M_A =130$ GeV.\label{tab:systdzmutauiia}}
\end{table}

\begin{table}[!htbp]
\begin{center}
\begin{tabular}{r|cccccc}
\hline
       Contribution & Signal           & diboson          &  \zmm       & \ztt                  &\ensuremath{t\bar{t}}  & QCD             \\ \hline
Muon identification &               4.0&               4.0&          4.0&                    4.0&                    4.0&                    0.0\\
Muon track reconstruction &               2.0&               2.0&          2.0&                    2.0&                    2.0&                    0.0\\
Tau identification  &               3.9&               0.0&          0.0&                    3.8&                    0.0&                    0.0\\
Tau track reconstruction &          1.4&               0.0&          0.0&                    1.4&                    0.0&                    0.0\\
Tau energy scale    &             2.6&               2.4&          2.7&                    2.5&                    2.3&                    0.0\\
Trigger             &               5.0&               5.0&          5.0&                    5.0&                    5.0&                    0.0\\
Signal acceptance                 &              4.60&              0.0&          0.0&                    0.0&                    0.0&                    0.0\\
MC cross sections   &              0.0&               5.0&          5.0&                    5.0&                    5.0&                    0.0\\
QCD                 &               0.0&               0.0&           0.0&                    0.0&                    0.0&                    22\\
Luminosity          &  6.1&               6.1&          6.1&                    6.1&                    6.1&                    0.0\\\hline
\end{tabular}
\end{center}
\caption{Percentage systematic uncertainties for each distribution in the D\O\ \mutau\ - (RunIIb) - combined across all three tau categories. Signal uncertainties are for $M_A$ = 130 GeV.\label{tab:systdzmutauiib}}
\end{table}

\begin{table}[!htbp]
\begin{center}
\begin{tabular}{r|cccccccc}
\hline
 Contribution          & Signal     & QCD         & \ensuremath{t\bar{t}}& \wlnu       &  diboson          & \zee              & \zmm        & \ztt  \\ \hline
Jet energy scale    &            2.0&          0.0&                   2.0&          2.0&                2.0&                2.0&  2.0 &             2.0\\
Electron identification&         2.0&          0.0&                   2.0&          2.0&                2.0&                2.0&  2.0 &             2.0\\
Muon identification &           0.4&          0.0&                  0.4&         0.4&               0.4&               0.4&  0.4 &            0.4\\
Vertex modelling     &            2.0&          0.0&                   2.0&          2.0&               2.0&                 2.0&  2.0 &             2.0\\
Trigger             &            4.0&          0.0&                   4.0&          4.0&                4.0&                4.0&  4.0 &             4.0\\
Signal acceptance   &            4.0&          0.0&                   0.0&          0.0&                0.0&                0.0&  0.0 &             0.0\\
MC cross sections   &            0.0&          0.0&                   5.0&          5.0&                5.0&                5.0&  5.0 &             5.0\\
QCD                 &            0.0&          7.0&                   0.0&          0.0&                0.0&                0.0&  0.0 &             0.0\\
Luminosity          &            6.1&          0.0&                   6.1&          6.1&                6.1&                6.1&  6.1 &             6.1\\\hline
\end{tabular}
\end{center}
\caption{Percentage systematic uncertainties each distribution in the D\O\ \emu\ analysis. Signal uncertainties are for $M_A$ = 130 GeV.\label{tab:systdzemu}}
\end{table}

\section{Combined Results} 

\begin{figure}[!htbp]
  \begin{center}
    \includegraphics[width=0.7\linewidth]{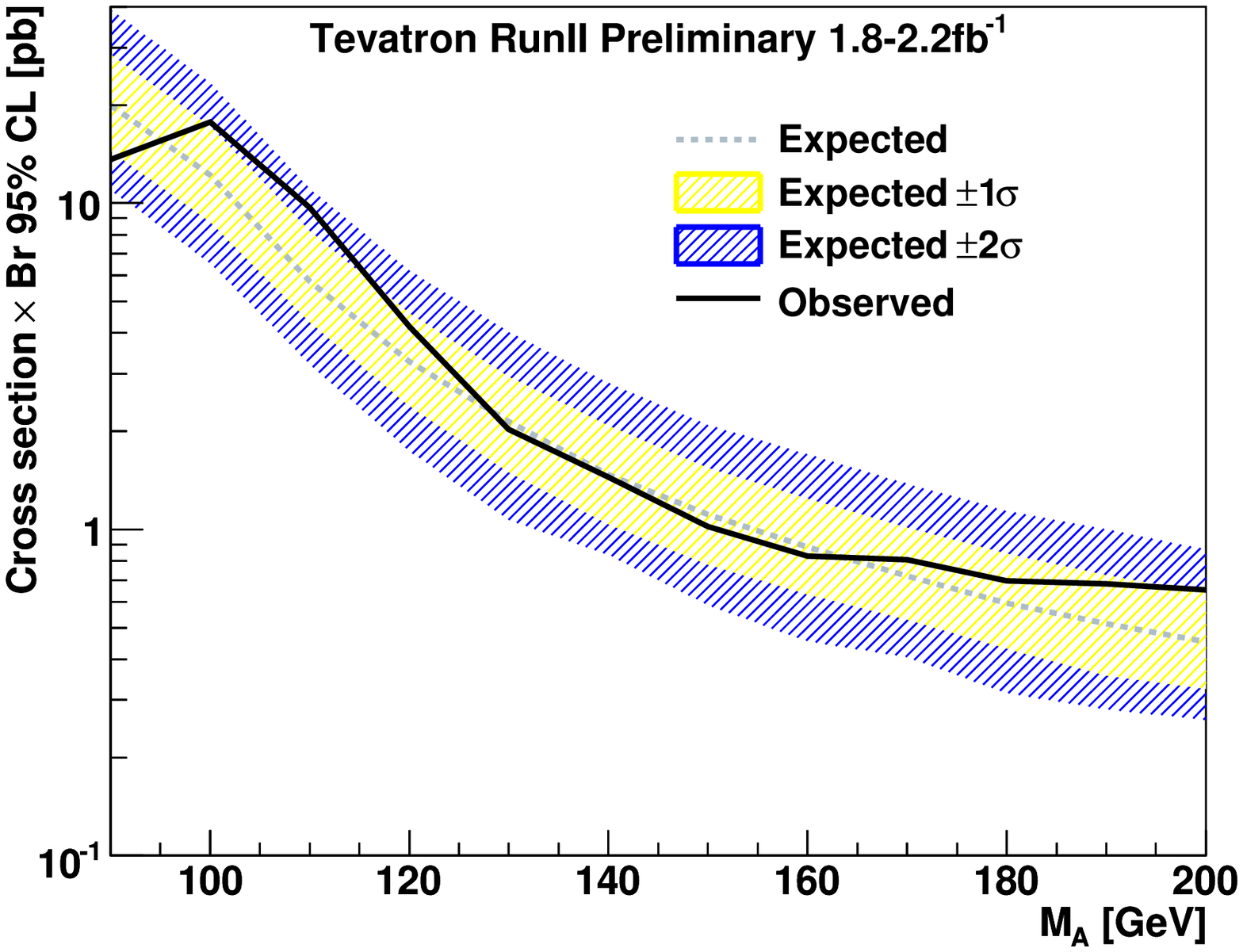}
\end{center}
  \caption{95\% Confidence limits on cross section $\times$ branching ratio. The solid black and dashed grey lines show the observed and expected limits respectively. The yellow and blue hatched bands around the expected limit show the 1 and 2$\sigma$ deviations from the expectation.\label{fig:xseclimit}}
\end{figure}

Tables \ref{tab:bayes} and \ref{tab:cls} give the 95\% confidence limits on the cross section $\times$ branching ratio for MSSM Higgs production and decay in the di-tau channel, using the two different approaches outlined above. Good agreement in the results for the two procedures is seen with variations at less than 10\%. The results are shown graphically in Figure \ref{fig:xseclimit}, using the $CL_s$ calculations. The observed limits are generally in good agreement with expectation with no evidence for significant excess for $90 < M_A< 200$~GeV.

\begin{table}
\begin{center}
\begin{tabular}{c|dddddd}
\hline
            & \mathrm{Observed} & \multicolumn{5}{c}{Expected Limits / pb}\\
Mass / GeV & \mathrm{Limits / pb} & -2 \ensuremath{\sigma} & -1\ensuremath{\sigma} & \mathrm{median} & +1\ensuremath{\sigma} & +2\ensuremath{\sigma} \\ \hline
 90&   12.6 &    10.6   &     14.6  &    20.2   &    28.7   &    39.2 \\
100&   17.4 &    6.89   &     9.52  &    13.2   &    18.6   &    25.1   \\
110&   9.89 &    3.30   &     4.45  &    6.18   &    8.63   &    11.6   \\
120&   4.22  &    1.77   &     2.38  &    3.40   &    4.80   &    6.69   \\
130&   1.94  &    1.16   &    1.55   &    2.16   &    2.92   &    3.93   \\
140&   1.41  &    0.805  &    1.08   &    1.55   &    2.10   &    2.80   \\
150&   1.00  &    0.615  &    0.795  &    1.13   &    1.56   &    2.14   \\
160&   0.832 &    0.495  &    0.655  &    0.895  &    1.26   &    1.71   \\
170&   0.771 &    0.405  &    0.525  &    0.745  &    1.07   &    1.41   \\
180&   0.647 &    0.325  &    0.445  &    0.615  &    0.865  &    1.65   \\
190&   0.628 &    0.295  &    0.385  &    0.545  &    0.765  &    1.04   \\
200&   0.629 &    0.255  &    0.335  &    0.475  &    0.665  &    0.915  \\ \hline
\end{tabular}
\end{center}
\caption{Combined Cross section $\times$ branching ratio limits using Bayes method\label{tab:bayes}.}
\end{table}

\begin{table}[!htbp]
\begin{center}
\begin{tabular}{c|dddddd}
\hline
 & \mathrm{Observed} & \multicolumn{5}{c}{Expected Limits / pb}\\
 Mass / GeV & \mathrm{Limits / pb} & -2\ensuremath{\sigma} & -1\ensuremath{\sigma} & \mathrm{median} & +1\ensuremath{\sigma} & +2\ensuremath{\sigma} \\ \hline
 90&    13.6 &    11.0 &   14.5  &    20.0 &    28.6 &    39.3\\
100&    17.7 &    6.55 &    8.77 &    12.2 &    17.2 &    23.4\\
110&    9.68 &    3.21 &    4.32 &    5.76 &    8.20 &    11.1\\
120&    4.19 &    1.74 &    2.39 &    3.26 &    4.60 &     6.22\\
130&    2.03 &    1.07 &    1.50 &    2.14 &    2.94 &     4.03\\
140&    1.45 &   0.837 &    1.04 &    1.47 &    2.08 &     2.81\\
150&    1.02 &   0.586 &   0.780 &    1.11 &    1.54 &     2.09\\
160&    0.829&   0.454 &   0.633 &   0.884 &    1.24 &     1.70\\
170&    0.807&   0.406 &   0.529 &   0.719 &    1.01 &     1.38\\
180&    0.697&   0.315 &   0.431 &   0.595 &   0.841 &     1.14\\
190&    0.681&   0.281 &   0.357 &   0.514 &   0.724 &     1.00\\
200&    0.653&   0.261 &   0.325 &   0.452 &   0.638 &     0.867\\ \hline
\end{tabular}
\end{center}
\caption{Combined cross section $\times$ branching ratio limits using $\mathrm{CL}_\mathrm{S}$\label{tab:cls}.}
\end{table}

\section{Interpretation within the MSSM}
Though at leading order the Higgs sector of the MSSM can be described with just two parameters, with higher order corrections comes a dependence on other model parameters. To interpret the exclusion within the MSSM these parameters are fixed in four benchmark scenarios~\cite{scenarios}. The four scenarios considered are defined in terms of: $M_{SUSY}$, the mass scale of squarks, $\mu$, the Higgs sector bilinear coupling, $M_2$, the gaugino mass term, $A_t$, the trilinear coupling of the stop sector, $A_b$, the trilinear coupling of the sbottom sector and $m_{\tilde{g}}$ the gluino mass term. The maximal-mixing, $m_h^{max}$, scenario is defined as:

\begin{eqnarray*}{}
  &M_{\mathrm{SUSY}} = 1 \mathrm{TeV}, \mu = 200 ~\mathrm{GeV}, M_2 = 200 ~\mathrm{GeV},\\
  &X_t = 2M_{\mathrm{SUSY}} \\
  &A_b = A_t, m_{\tilde{g}} = 0.8M_{\mathrm{SUSY}}.
\end{eqnarray*}

and the no-mixing scenario - with vanishing mixing in the stop sector and a higher SUSY mass scale to avoid the LEP Higgs bounds:

\begin{eqnarray*}{}
  &M_{\mathrm{SUSY}} = 2 \mathrm{TeV}, \mu = 200 ~\mathrm{GeV}, M_2 = 200 ~\mathrm{GeV}, \\
  &X_t = 0 , A_b = A_t, m_{\tilde{g}} = 0.8M_{\mathrm{SUSY}}.
\end{eqnarray*}

Four scenarios are constructed from these two by the consideration of both + and - signs for $\mu$. 

\begin{figure}[!htbp]
  \begin{center}
    \begin{tabular}{cc}
      \includegraphics[width=0.45\textwidth]{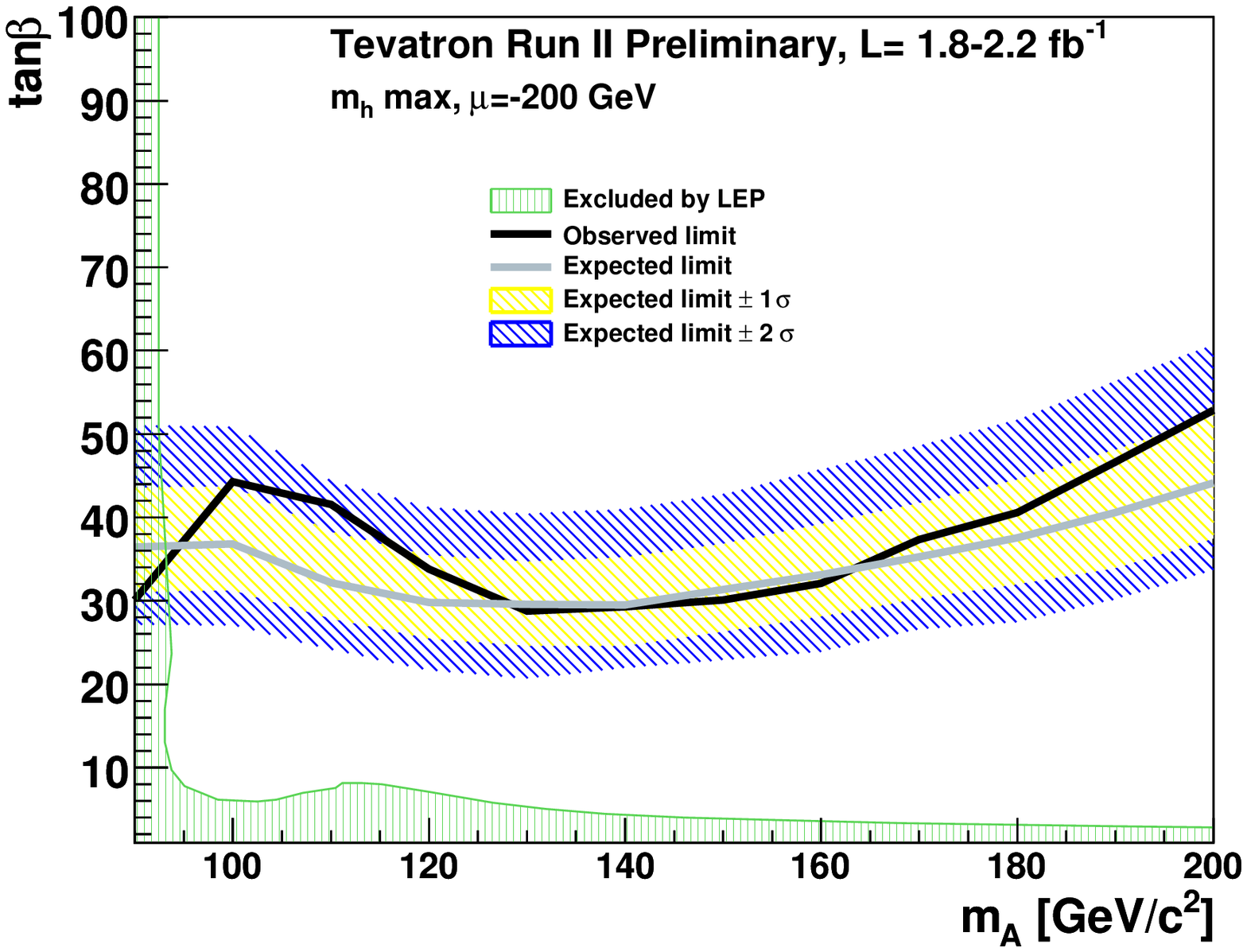} & 
      \includegraphics[width=0.45\textwidth]{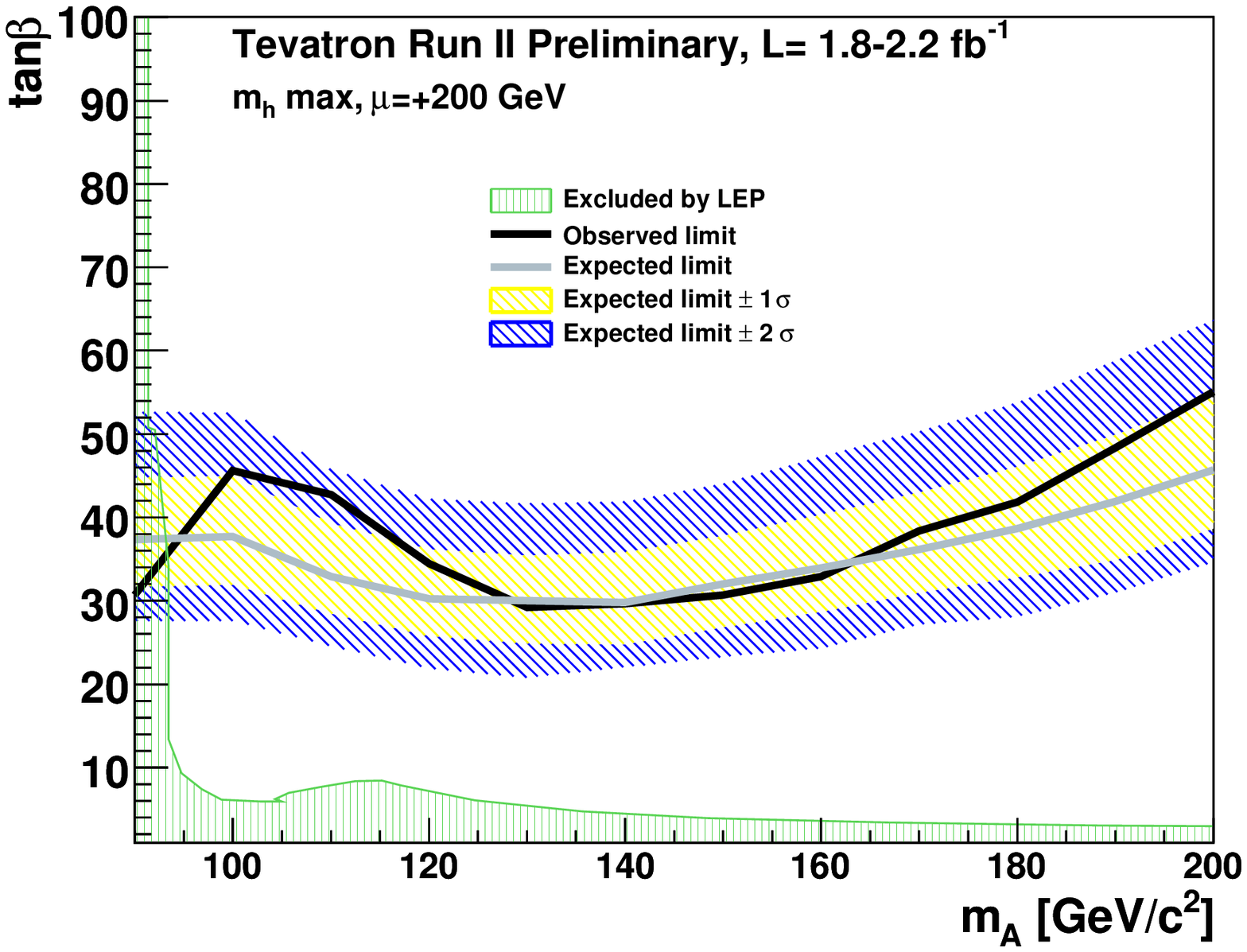} \\
      \includegraphics[width=0.45\textwidth]{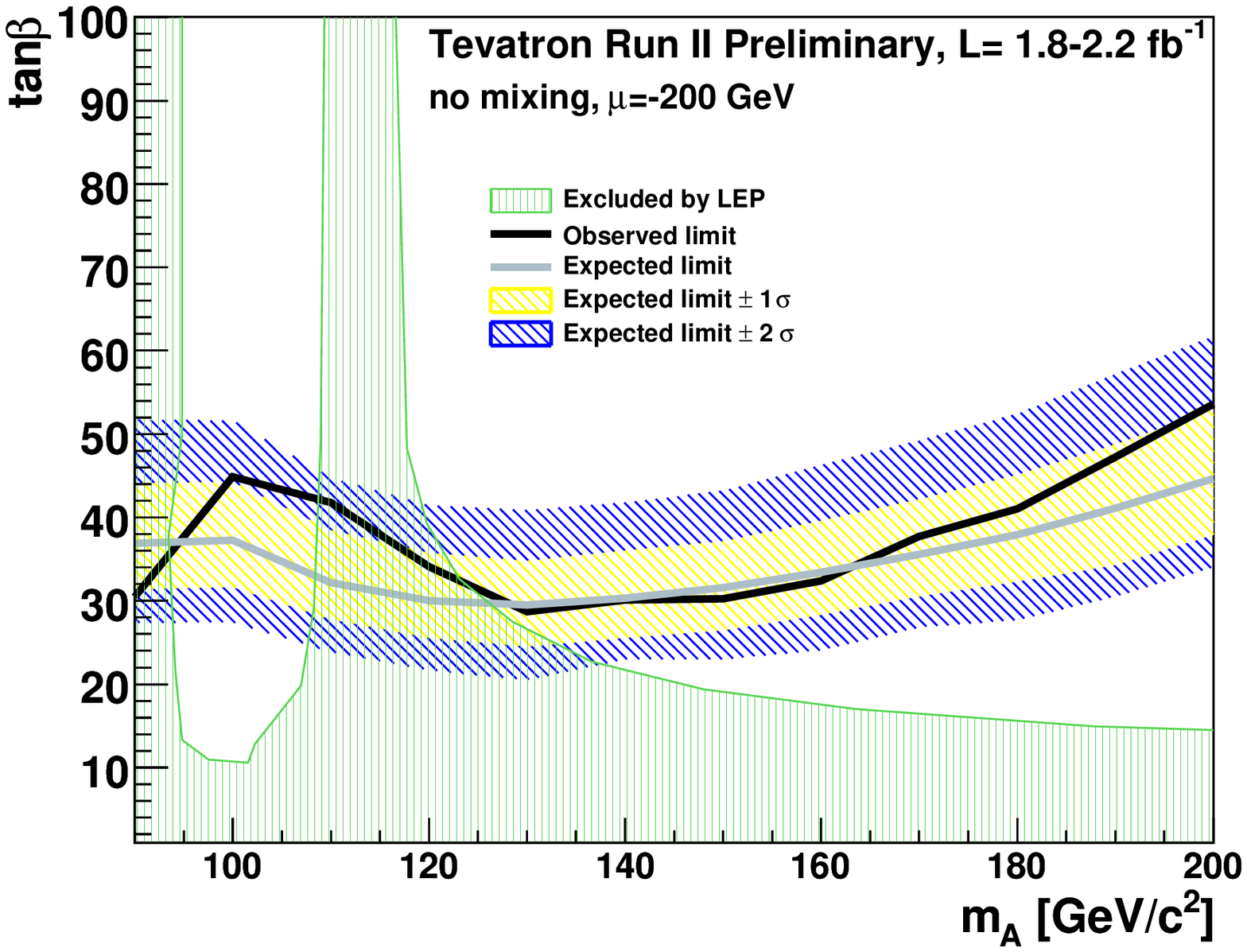} & 
      \includegraphics[width=0.45\textwidth]{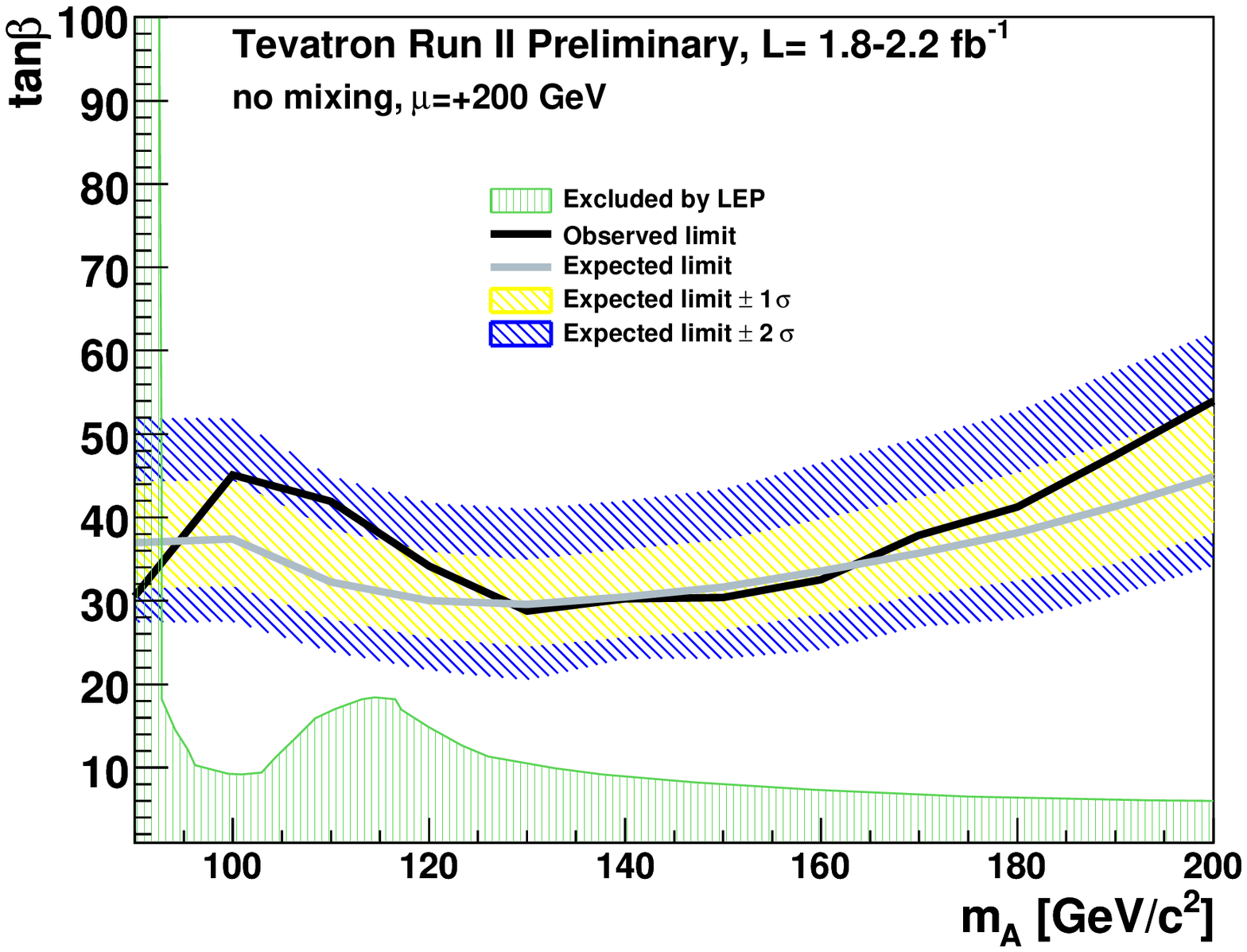} \\
      \end{tabular}
\end{center}
  \caption{95\% Confidence limits in the \tanb-M$_{\mathrm{A}}$ plane for the 4 benchmark scenarios: maximal mixing (top) and no mixing (bottom) for $\mu <0$ (left) and $\mu > 0$ (right). The black line denotes the observed limit, the grey line the expected limit and the hatched yellow and blue regions denote the $\pm$1 and 2 $\sigma$ bands around the expectation. The shaded light-green area shows the limits from LEP. \label{fig:tanblimits}}
\end{figure}

Tables \ref{tab:mixplus}, \ref{tab:mixminus}, \ref{tab:nomixplus}, and \ref{tab:nomixminus} give the observed and median expected 95\% confidence limits on \tanb~ for the tested mass hypotheses for the four different benchmark scenarios considered. This is shown graphically in Figure \ref{fig:tanblimits}. 

\begin{table}[htb]
\begin{center}
\begin{tabular}{c|cccccc}
\hline
 & Observed & \multicolumn{5}{c}{Expected Limits / pb}\\
 $M_{A}$ GeV & Limits & -2$\sigma$ & -1$\sigma$ & median & +1$\sigma$ & +2$\sigma$ \\ \hline
 90&  30     &   27    &   31    &  36     &  44     &  51    \\
100&  44     &   27    &   31    &  37     &  44     &  51    \\
110&  42     &   24    &   28    &  32     &  38     &  44    \\
120&  34     &   22    &   25    &  30     &  35     &  41     \\
130&  29     &   21    &   25    &  30     &  35     &  40     \\
140&  29     &   22    &   25    &  29     &  35     &  41     \\
150&  30     &   23    &   26    &  31     &  37     &  43     \\
160&  32     &   24    &   28    &  33     &  39     &  46     \\
170&  37     &   27    &   30    &  35     &  42     &  49     \\
180&  41     &   27    &   32    &  38     &  45     &  52     \\
190&  47     &   30    &   34    &  41     &  48     &  56     \\
200&  53     &   34    &   38    &  44     &  52     &  61      \\\hline
\end{tabular}
\end{center}
\caption{Combined 95\% confidence limits on tan$\beta$ for each mass hypothesis in the  $m_h$ max and negative $\mu$ scenario.\label{tab:mixminus}}
\end{table}

\begin{table}[htb]
\begin{center}
\begin{tabular}{c|cccccc}
\hline
 & Observed & \multicolumn{5}{c}{Expected Limits / pb}\\
 $M_{A}$ GeV & Limits & -2$\sigma$ & -1$\sigma$ & median & +1$\sigma$ & +2$\sigma$ \\ \hline
 90&  31     &   28    &   32    &  37     &  45     &  53    \\
100&  46     &   28    &   32    &  38     &  45     &  53    \\
110&  43     &   25    &   28    &  33     &  40     &  46    \\
120&  34     &   22    &   26    &  30     &  36     &  42     \\
130&  29     &   21    &   25    &  30     &  36     &  42     \\
140&  30     &   22    &   25    &  30     &  36     &  42     \\
150&  31     &   23    &   27    &  32     &  38     &  44     \\
160&  33     &   24    &   29    &  34     &  40     &  47     \\
170&  38     &   27    &   31    &  36     &  43     &  50     \\
180&  42     &   28    &   33    &  39     &  46     &  54     \\
190&  48     &   31    &   35    &  42     &  50     &  59     \\
200&  55     &   35    &   39    &  46     &  54     &  64      \\\hline
\end{tabular}
\end{center}
\caption{Combined 95\% confidence limits on tan$\beta$ for each mass hypothesis in the  $m_h$ max and positive $\mu$ scenario.\label{tab:mixplus}}
\end{table}

\begin{table}[htb]
\begin{center}
\begin{tabular}{c|cccccc}
\hline
 & Observed & \multicolumn{5}{c}{Expected Limits / pb}\\
 $M_{A}$ GeV &  Limits  & -2$ \sigma$ & -1$ \sigma$ &  median  & +1$ \sigma$ & +2$ \sigma$ \\ \hline
 90&  30     &   27    &   31    &  37     &  44     &  52    \\
100&  45     &   27    &   32    &  37     &  44     &  52    \\
110&  42     &   24    &   28    &  32     &  38     &  45    \\
120&  34     &   22    &   26    &  30     &  36     &  41     \\
130&  29     &   20    &   25    &  30     &  35     &  41     \\
140&  30     &   23    &   26    &  30     &  36     &  42     \\
150&  30     &   23    &   26    &  32     &  37     &  43     \\
160&  32     &   24    &   28    &  33     &  40     &  46     \\
170&  38     &   27    &   31    &  36     &  42     &  49     \\
180&  41     &   28    &   32    &  38     &  45     &  52     \\
190&  47     &   30    &   34    &  41     &  49     &  57     \\
200&  54     &   34    &   38    &  45     &  53     &  62      \\\hline
\end{tabular}
\end{center}
\caption{Combined 95\% confidence limits on tan$\beta$ for each mass hypothesis in the no-mixing and negative $\mu$ scenario.\label{tab:nomixminus}}
\end{table}

\begin{table}[htb]
\begin{center}
\begin{tabular}{c|cccccc}
\hline
 & {Observed} & \multicolumn{5}{c}{Expected Limits / pb}\\
 $M_{A}$ GeV &  Limits  & -2$ \sigma$ & -1$ \sigma$ & median & +1$ \sigma$ & +2$ \sigma$ \\ \hline
 90&  31     &   27    &   31    &  37     &  44     &  52     \\
100&  45     &   27    &   32    &  37     &  44     &  52    \\
110&  42     &   24    &   28    &  32     &  39     &  45    \\
120&  34     &   22    &   26    &  30     &  36     &  42     \\
130&  29     &   20    &   25    &  30     &  35     &  41     \\
140&  30     &   23    &   26    &  30     &  36     &  42     \\
150&  30     &   23    &   27    &  32     &  37     &  43     \\
160&  33     &   24    &   28    &  34     &  40     &  47     \\
170&  38     &   27    &   31    &  36     &  42     &  50     \\
180&  41     &   28    &   32    &  38     &  45     &  53     \\
190&  47     &   31    &   35    &  41     &  49     &  58     \\
200&  54     &   34    &   38    &  45     &  53     &  62      \\\hline
\end{tabular}
\end{center}
\caption{Combined 95\% confidence limits on tan$\beta$ for each mass hypothesis in the no-mixing and positive $\mu$ scenario.\label{tab:nomixplus}}
\end{table}

In this preliminary result the signal cross sections and branching fractions within each scenario have been calculated using {\sc feynhiggs}~\cite{feynhiggs}
- with $gg\rightarrow H$ production from \cite{Georgi:1977gs, Djouadi:1991tka, Dawson:1990zj, Spira:1995rr, Bonciani:2007ex, Aglietti:2006tp, Harlander:2002wh, Anastasiou:2002yz, Ravindran:2003um, Catani:2003zt, Marzani:2008az} and SM $bb\rightarrow H$ production from \cite{harlander:2003} and references therein and MRST2002 NNLO PDFs~\cite{mrst2002} - with no theoretical uncertainties considered. Tan$\beta$ dependent width effects have not been included, though in the region of the tan$\beta$-$M_{A}$ plane where limits have been set these are not expected to strongly affect the limit~\cite{p17htt}.

This combination of Tevatron results from CDF and D0 in the \Att\ channel sets the most stringent limits to date on the search for MSSM Higgs in that final state.

\end{document}